\documentclass[%
reprint,
amsmath,amssymb,
aps,
superscriptaddress,
prd
]{revtex4-1}

\usepackage{graphicx}
\usepackage{dcolumn}
\usepackage{bm}
\usepackage{hyperref}

\usepackage[capitalise]{cleveref}
\usepackage{comment}

\begin{document}

	\title{Investigating jet induced identified hadron productions from the relative transverse activity classifier in pp collision at LHC}
	
	\author{Yuhao Peng}
	\affiliation{School of Mathematics and Physics, China University of
		Geosciences (Wuhan), Wuhan 430074, China}

	\author{Yan Wu}
	\affiliation{School of Mathematics and Physics, China University of
	Geosciences (Wuhan), Wuhan 430074, China}

	\author{Xinye Peng}
	\affiliation{School of Mathematics and Physics, China University of
		Geosciences (Wuhan), Wuhan 430074, China}
	\affiliation{Key Laboratory of Quark and Lepton Physics (MOE) and Institute
		of Particle Physics, Central China Normal University, Wuhan 430079, China}

	\author{Zhongbao Yin}
	\affiliation{Key Laboratory of Quark and Lepton Physics (MOE) and Institute
	of Particle Physics, Central China Normal University, Wuhan 430079, China}
  
	\author{Liang Zheng}\email{zhengliang@cug.edu.cn}
	\affiliation{School of Mathematics and Physics, China University of
		Geosciences (Wuhan), Wuhan 430074, China}
	\affiliation{Shanghai Research Center for Theoretical Nuclear Physics, NSFC and Fudan University, Shanghai 200438, China}
	\affiliation{Key Laboratory of Quark and Lepton Physics (MOE) and Institute
	of Particle Physics, Central China Normal University, Wuhan 430079, China}	
	
	\date{\today}
	
	\begin{abstract}
In this work, we systematically investigate the jet associated identified hadron productions of pions, kaons, and protons based on the event topological separation method in proton proton (pp) collisions at $\sqrt{s}=13$ TeV employing the AMPT model with PYTHIA8 initial conditions. Enabling relative transverse event activity classifier $R_T$, we analyze the transverse momentum ($p_T$) spectra and particle ratios in the jet aligned toward region and the underlying event (UE) contributions sensitive transverse region varying with $R_T$. The results indicate that the AMPT model, incorporating both partonic and hadronic final-state interactions, provides a satisfactory description to the experimental data of the $p_T$ differential yields, particle ratios and average transverse momentum of identified particles in both the toward and transverse regions across different event activity classes. By subtracting the UE contribution from the toward region using the transverse region hadron yield, we introduce the in-jet hadron productions to analyze the modifications to jet itself.  We find that the in-jet baryon to meson ratios reveal a sensitive dependence on $R_T$ driven by final state interactions, leading to a unique crossing pattern at intermediate $p_T$. This behavior can be regarded as a novel signature for probing jet medium interactions and energy loss effects in small collision systems. 

	\end{abstract}
	

	\maketitle
	
	
	\section{Introduction}
	\label{sec:level1}

In heavy ion experiments, collectivity signatures observed during high energy nuclear collisions provide compelling evidence for the creation of quark gluon plasma (QGP) matter, which exhibits a collective, fluid like expansion throughout its evolution~\cite{Harris:1996zx,Elfner:2022iae,Busza:2018rrf,Chen:2024aom,Shou:2024uga}.  
In recent years, it has been revealed by the experiments at the Large Hadron Collider (LHC) that flow like signals remarkably similar to those seen in heavy ion collisions can also be found in systems as small as proton proton (pp) collisions, which were traditionally assumed to provide a baseline free of collective effects~\cite{Nagle:2018nvi,Adolfsson:2020dhm,Noronha:2024dtq,Grosse-Oetringhaus:2024bwr}. Particle production in these smaller systems has often been modeled based on incoherent superpositions of multiple parton interaction (MPI), in which the created partonic degrees of freedom are expected to free stream and fragment in vacuum without interacting with a medium~\cite{Sjostrand:2004pf}. The unexpected collective behavior observed in pp collisions suggests that the underlying event (UE), which encompasses all activity beyond the primary hard scatter, may act like an evolving medium with high energy densities to interact the initial partons rather than a simple uncorrelated background~\cite{Gale:2013da,dEnterria:2010xip,Adolfsson:2020dhm,Altmann:2024icx}.

Despite the presence of sizable anisotropic flow, characterized by long range azimuthal correlations and the mass dependent hardening of transverse momentum spectra, a clear observation of parton energy loss or jet quenching signals remains elusive in small systems~\cite{CMS:2014qvs,ALICE:2017svf,ALICE:2023ama,Harris:2023tti}. The absence of definitive evidence for jet quenching raises fundamental questions about the nature of the medium created in small systems and the mechanisms governing energy loss in these environments~\cite{CMS:2025kzg,Grosse-Oetringhaus:2024bwr}. It is hypothesized that parton energy loss effect in small systems might be too subtle to be detected with current experimental precision~\cite{ALICE:2017svf,ATLAS:2022iyq}, or that systematic uncertainties due to selection bias in measurements remain significant~\cite{ALICE:2017svf}. A primary challenge lies in how to isolate a subtle jet modification signal from the large and fluctuating UE background.
To address this discrepancy, advanced experimental techniques have been proposed to detect subtle modifications to the internal structure of fully reconstructed jets, such as groomed jet momentum fraction~\cite{CMS:2018fof,JETSCAPE:2023hqn,Chien:2016led,Milhano:2017nzm,Duan:2025wsy} and energy energy correlators~\cite{Gao:2019ojf,Moult:2025nhu}. These measurements are expected to precisely map the angular distribution of energy within the jet, thus offering sensitive probes to the potential jet medium interactions. However, full jet reconstruction and grooming analyses are typically limited to high $p_T$ jets, where  background contamination is manageable. These approaches lose sensitivity in the soft to intermediate $p_T$ regime, where the interplay between jet fragments and UE particles becomes significant. In contrast, the event topology based background subtraction method is expected to provide a complementary framework that does not rely on explicit jet finding process and allows us to study the jet induced particle productions including the soft fragmentation region with minimal selection bias.

Unraveling the parton energy loss phenomena in small systems requires distinguishing jet particles from UE contributions on an event-by-event basis.
While selecting events based on global observables like charged particle multiplicity is a common approach, this technique introduces a strong bias towards enhanced UE activity from multiple hard scatterings in high multiplicity events, thereby entangling jet related effects with UE dynamics~\cite{ALICE:2022qxg,ALICE:2018pal,Mendez:2025dqz}.
A more robust and less biased approach is to categorize events using topological classifiers such as the relative transverse event activity $R_T$. Calculated as the ratio of the charged multiplicity in the transverse region to its average value, $R_T$ allows for a more differential study of particle production with the sensitivity to separate UE activity and hard scattering effects.~\cite{Martin:2016igp,ALICE:2023yuk,ALICE:2019mmy,Bencedi:2021tst}. By classifying events according to $R_T$, one can effectively probe particle production across different UE environments with reduced bias from hard scattering. Correlating the event activity observables with collectivity measurements can be of great interest to understand the origin of the flow signal in small systems~\cite{Bencedi:2020qpi,ALICE:2023yuk,Prasad:2024gqq,Prasad:2025yfj}.
Furthermore, by leveraging the expectation that the toward region is more dominant by jet fragmentation, the differences between the toward and transverse regions can be used to directly probe jet fragmentation itself and expose jet medium interaction effects~\cite{ALICE:2022ugx,Feng:2024tmc}. This subtraction method, based on the azimuthal correlation technique, allows us to isolate the contribution of particles associated with the jet from the dominant UE background in a way that complements full jet reconstruction analyses. Compared to previous measurements employing full jet reconstructions to directly investigate the production of identified hadrons within jets~\cite{ALICE:2022ecr,Cui:2022puv}, an advantage of this approach is its sensitivity to jet modification effects even at lower $p_T$ region where full jet reconstruction may be less efficient~\cite{ALICE:2024iqc,ALICE:2024zxp}.

In this paper, we present a systematic investigation into the production of $\pi$, $K$, $p$, in pp collisions at $\sqrt{s} = 13$ TeV, utilizing the transverse activity classifier. We employ the AMPT model~\cite{Lin:2004en,Lin:2021mdn}, which includes both parton and hadron final state interactions, to analyze the $p_T$ spectra and particle ratios in the toward and transverse regions. Crucially, we define the in-jet yield as the difference of particle productions between the toward and transverse regions to understand the modifications to the jet objects induced by the interactions with the medium. By studying these observables, we aim to provide insights into the hadronization mechanisms and the nature of collectivity like phenomena in high multiplicity pp collisions, offering valuable constraints to understand the interplay between jet fragmentation and the underlying event driven medium evolutions in small systems. The structure of this paper is as follows: Sect.~\ref{sec:formalism} describes the methodology, including the AMPT model setup and the definition of the $R_T$ classifier. Section~\ref{sec:results} presents the results on identified hadron $p_T$ spectra and particle ratios as a function of $R_T$ in different topological regions. Finally, Sect.~\ref{sec:summary} provides a summary and discussion of the findings.

\section{Method}\label{sec:formalism}
\subsection{The AMPT model}
    In this study, we have carried out the research using the AMPT model based on PYTHIA8 initial conditions to explore the impacts of final state parton and hadron interactions on the hadron productions in different topological event regions. The string melting AMPT model consists of four major ingredients: fluctuating initial conditions, final state parton cascades, coalescence hadronization, and final state hadronic cascades. The event by event fluctuating initial conditions for the subsequent evolution stage are generated by PYTHIA8 together with spatial fluctuations at the sub-nucleon level~\cite{Zheng:2021jrr,Zhang:2025pqu}. The resultant quark system may experience the parton rescattering stage with the microscopic two-body scattering process implemented in Zhang’s Parton Cascade (ZPC) model~\cite{Zhang:1997ej}. In this work, the value of this parton parton scattering cross section is set to $\sigma=0.15\mathrm{mb}$ which gives a satisfactory description to the elliptic flow measurements in pp collisions~\cite{Zheng:2024xyv}. When the partons cease interactions, an improved spatial coalescence model has been applied to convert the freeze-out partons into hadrons~\cite{He:2017tla}. An overall coalescence parameter $r_{BM}$ has been introduced to determine the relative probability for a quark to become a meson or a baryon in this model. To describe the hadron yield for particles with different quark flavors, we follow the prescriptions in Refs.~\cite{Shao:2020sqr,Zheng:2019alz,Zhang:2025pqu} and adjust the coalescence parameter for each flavor sector. For non-strange light flavor quark clusters, we have the coalescence parameter $r_{BM}=0.53$. If a strange quark or a charm quark is involved in the coalescence process, the value of this parameter has been changed to $r_{BM}^s=0.9$ and $r_{BM}^c=1.4$, respectively. The hadrons after coalescence may undergo further hadronic rescatterings implemented in the extended relativistic transport model (ART)~\cite{Li:1995pra}. 
    
    In this work, we turn on the parton and hadron final state transport mechanism in a step by step way to explore the effects developed in different evolution stages. When both parton and hadron rescattering are disabled by setting the partonic cross section to 0 mb and turning off the ART, the results are labeled as ``0 mb w/o ART'', whose behavior should be similar to the pure PYTHIA string fragmentation predictions without any collective effect. If the parton rescattering stage is enabled with $\sigma=0.15\mathrm{mb}$ while hadron rescatterings are excluded, it is denoted as the ``0.15 mb w/o ART'', representing the case with only partonic final state interactions. When both final state parton and hadron rescattering effects are included, the results are indicated as ``0.15 mb w/ ART'' (all final-state interaction), in which the evolving system experiences the entire partonic and hadronic evolutions.  
    
    
\begin{figure}[htbp!]
	\begin{center}
		\includegraphics[width=0.5\textwidth]{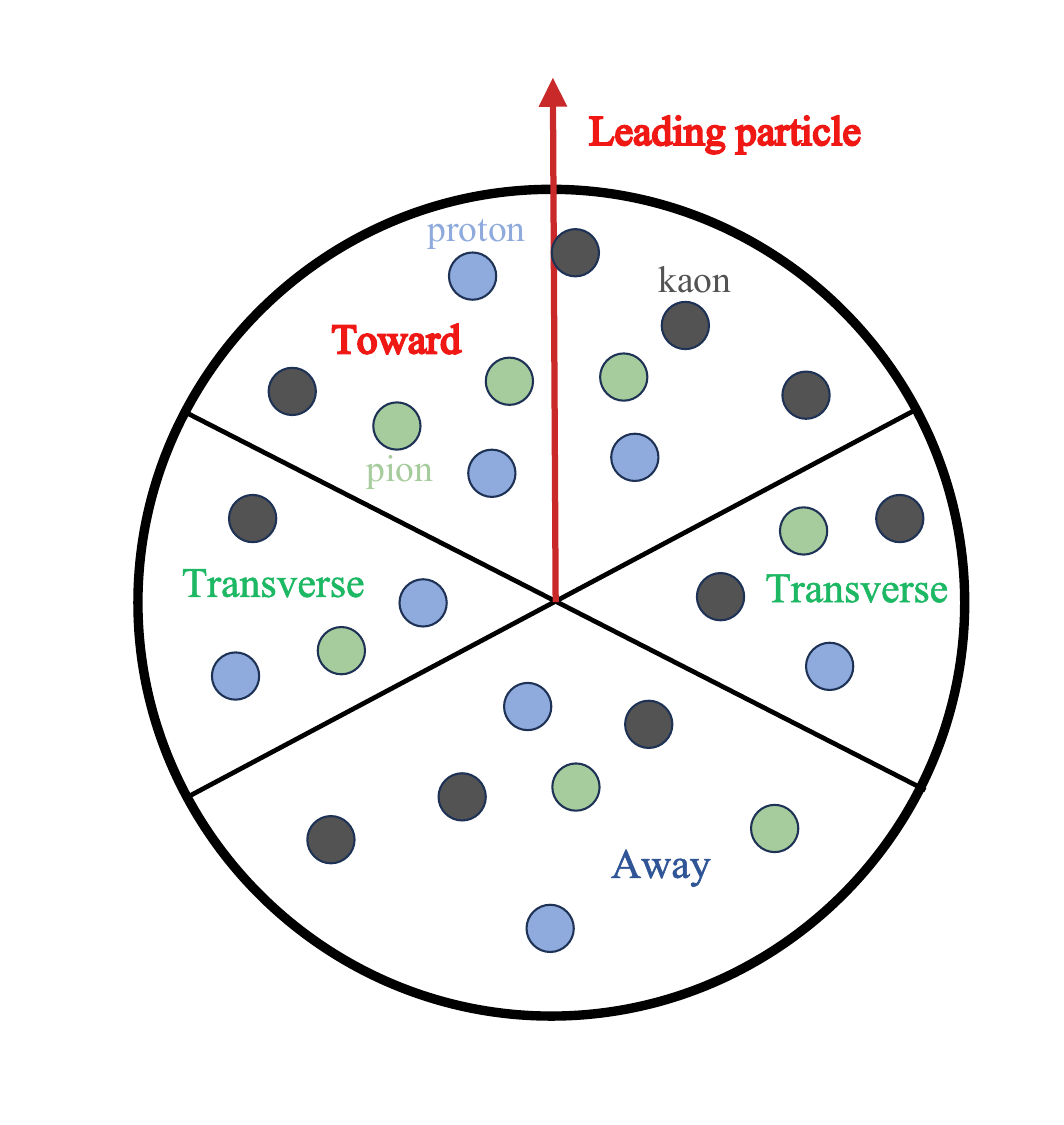}    
		\caption{$\pi$, K, p production in different event topological regions relative to the leading particle: $|\Delta\phi|<60^\circ$ (Toward), $|\Delta\phi|>120^{\circ}$ (Away), and $60^\circ<|\Delta\phi|<120^\circ$ (Transverse), with $\Delta\phi$ being the azimuthal angle of the emitted particle relative to that of the leading particle.       
		} \label{fig:region}
	\end{center}
\end{figure}    
    
\subsection{Event activity and in-jet production}\label{sec:rt_def}
	Understanding the event structure of pp collisions at high energies requires a careful separation of the hard scattering component from softer contributions associated with the underlying event. A widely used approach to achieve this is through the use of event topological regions, defined based on the azimuthal angle difference $\Delta\varphi = \varphi^{\mathrm{assoc}} - \varphi^{\mathrm{leading}}$ between the highest transverse momentum particle or jet and other associate particles in the event~\cite{CDF:2001onq, ALICE:2023yuk}. The azimuthal plane can be divided into three different topological regions as shown in Fig.~\ref{fig:region}: the toward region ($|\Delta\varphi| < 60^\circ$), aligned with the leading particle; the away region ($|\Delta\varphi| \geq 120^\circ$), opposite to the leading particle, typically containing the recoil jet; and the transverse region ($60^\circ \le |\Delta\varphi| < 120^\circ$), which lies perpendicular to the leading object direction. This approach allows for differential analysis to isolate jet activity from contributions arising due to softer interactions. The transverse region serves as a sensitive probe of the UE, which includes beam-beam remnants, initial and final-state radiation, and is expected to be less influenced by the leading jet~\cite{Sjostrand:1987su}. Therefore, the relative transverse activity (\( R_T \))~\cite{Martin:2016igp} built from the transverse region has been proposed to classify events and gain insight into the modifications to the hadron $p_T$ spectra
	
%

    \begin{equation}
    R_T = \frac{N_T}{\langle N_T\rangle},
    \end{equation}
    where $N_T$ is the charged particle multiplicity in the transverse region and $\langle N_T\rangle$ is its event ensemble average. $R_T$ provides a normalized measure of UE activity, offering a handle on MPI dynamics that scale with event multiplicity. By construction, $R_T$ separates events with higher than average UE from those with lower than average UE, irrespective of center of mass energy. This allows probing if jet dominated events at very low UE activity exhibit particle ratios and spectra consistent with fragmentation models following the jet universality, and conversely, if events with high UE activity show clear signs of flow or other collective effects. Investigations of identified hadron production in these topological regions have revealed clear dependencies on both $R_T$ and particle species. 

To further disentangle jet related production from UE effects, in-jet production is defined through a differential method as the difference in particle yield between the toward and transverse regions as follows: 
    \begin{equation}
     \frac{d^2 N^{\mathrm{In-Jet}}}{dp_{T}dy}=\frac{d^2 N^{\mathrm{Toward}}}{dp_{T}dy}-\frac{d^2 N^{\mathrm{Transverse}}}{dp_{T}dy}
    \end{equation}
This subtraction statistically reduces the influence of MPI and other soft processes that are expected to contribute equally or similarly across topological regions, thereby enhancing the signal due to genuine jet fragmentation and hard scatterings. The average transverse momentum for in-jet hadrons is calculated based on the subtracted yields between the toward and transverse regions
\begin{equation}
\langle p_T \rangle ^{\mathrm{In-Jet}} = \frac{\int p_T \frac{d^2 N^{\mathrm{In-Jet}}}{dp_{T}dy}dp_T} { \int \frac{d^2 N^{\mathrm{In-Jet}}}{dp_{T}dy}dp_T }.
\end{equation}
This definition ensures that the contribution from the UE is removed consistently from both the numerator and denominator, providing a measure of the mean transverse momentum of hadrons associated specifically with jet activity. The same $p_T$ range and rapidity acceptance as used in the yield analysis are applied here to maintain consistency across observables.

Analyzing in-jet particle yields and ratios can offer insights into the hadronization mechanisms within jets and possible medium modifications even in small systems~\cite{ALICE:2022ugx,Feng:2024tmc}. 
By comparing the in-jet particle composition across different event activity classes, one can investigate whether high UE activity environments modify the jet fragmentation process itself. If observed, such modifications may suggest a degree of coherence or interplay between the jet and the medium like environment in pp events, indicating QGP like behavior even in small systems.

\begin{figure*}[htbp!]
	\begin{center}
		\includegraphics[width=1.0\textwidth]{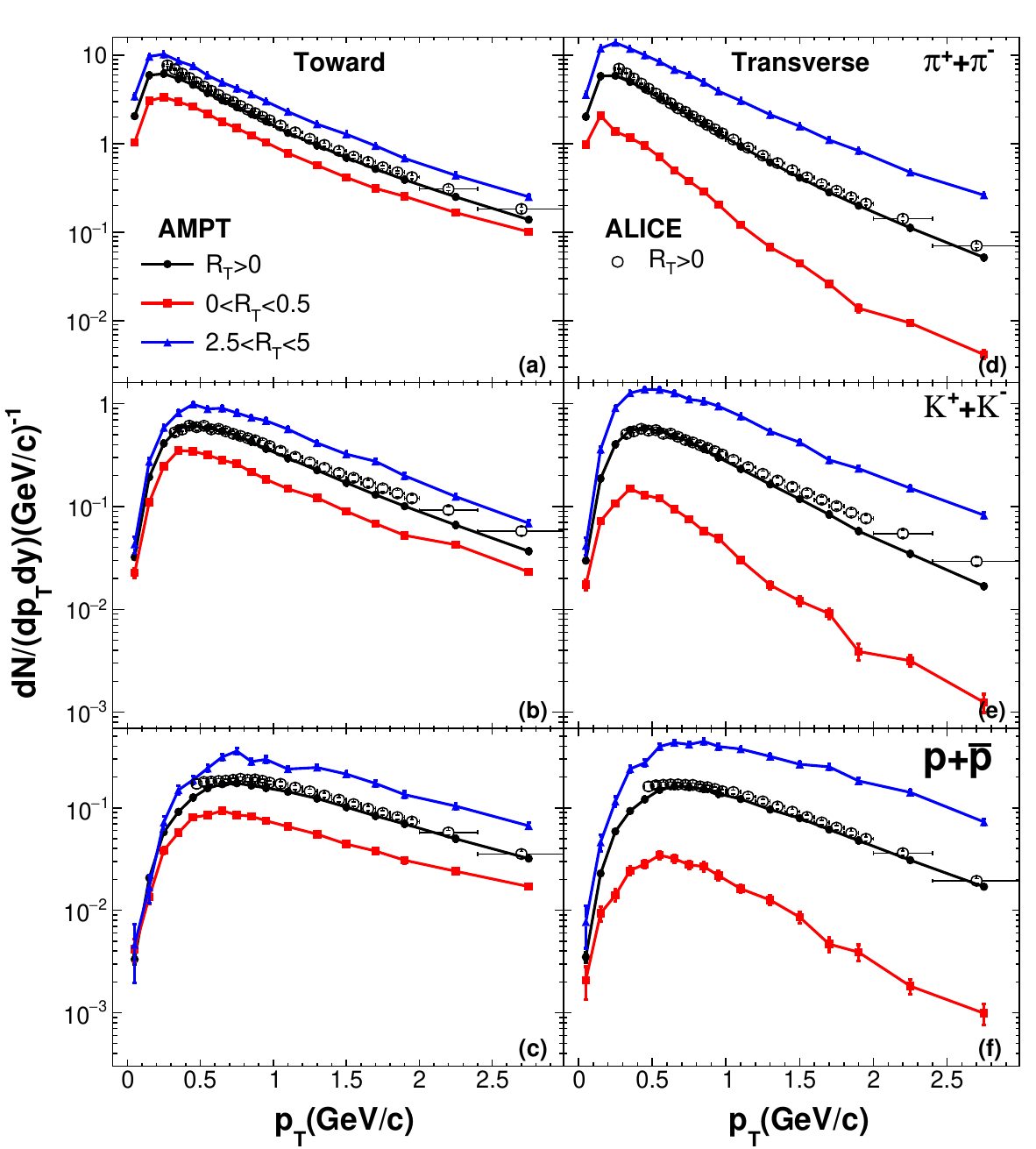}
		\caption{Transverse momentum spectra of $\pi^{\pm}$, $K^{\pm}$, and $p(\bar{p})$ in toward (left column) and transverse (right column) regions. The AMPT model calculations for $R_T$ integrated events, $0<R_T<0.5$ and $2.5<R_T<5$ are shown in the black lines, red lines and blue lines, respectively. ALICE data for $R_T$ integrated events taken from Ref.~\cite{ALICE:2023yuk} are indicated by the open markers for comparison.} \label{fig:pi_pt}
	\end{center}
\end{figure*}

\begin{figure}[htbp!]
	\begin{center}
		\includegraphics[width=0.45\textwidth]{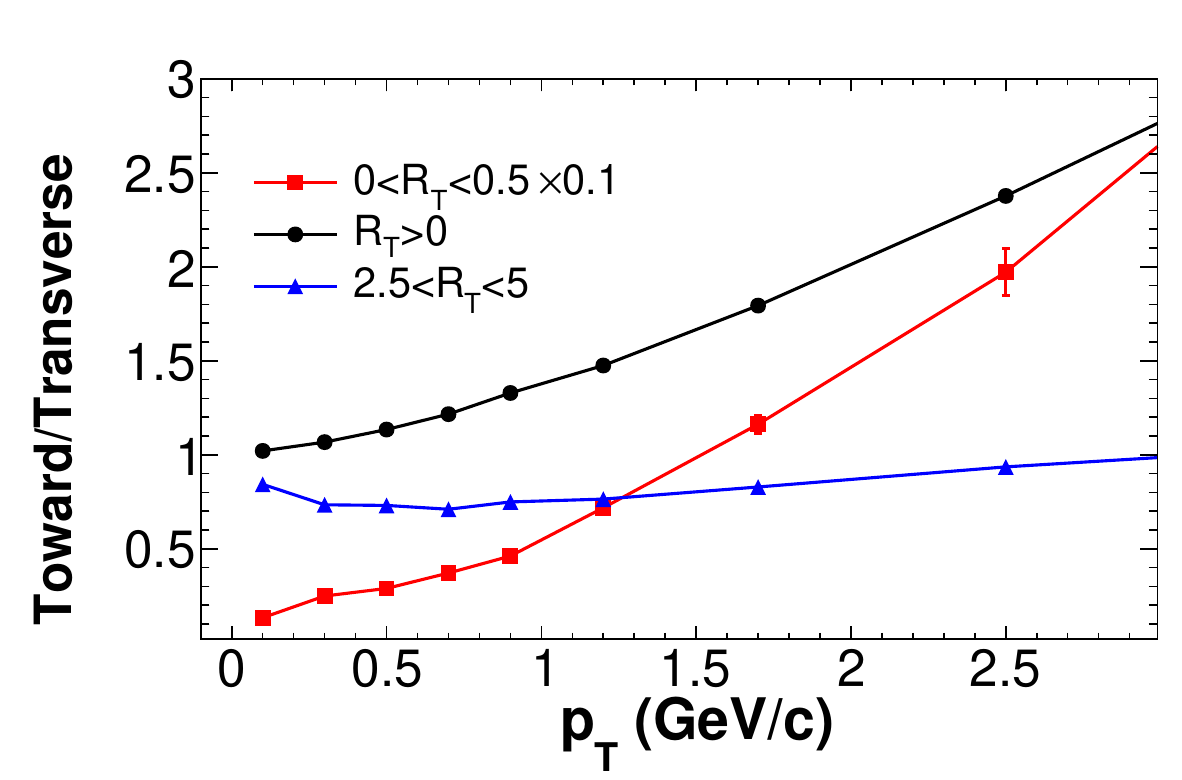}
		\caption{Ratio of the transverse-momentum spectra in the toward region to those in the transverse region for different $R_T$ event classes. The black, red and blue lines represent the results for the all events, $0<R_T<0.5$ and $2.5<R_T<5$, respectively. The $0<R_T<0.5$ results have been scaled by a factor 0.1 to allow comparison on the same vertical scale. } \label{fig:tow_to_trans_ratio}
	\end{center}
\end{figure}

\section{Results}\label{sec:results}
    
In this study, we examine the production of identified hadrons in pp collisions at $\sqrt{s}=13$ TeV across various topological regions. The analyzed events are required to contain a leading charged hadron with $p_T > 5$ GeV/$c$. Both the leading charged hadrons and associated hadrons are restricted to the mid-rapidity range $|\eta| < 0.8$. The event activity $R_T$ is calculated using the charged associate particle numbers in the transverse region, following the definition in Sec.~\ref{sec:rt_def}.

%
   In Fig.~\ref{fig:pi_pt}, we present the transverse momentum distributions of $\pi^{\pm}$, $K^{\pm}$, and $p(\bar{p})$ in toward and transverse regions from different event activity classes. The results in the toward and transverse regions are shown in the left and right columns, respectively. The solid lines indicate the AMPT calculations with both partonic and hadronic final state cascade effects included. Experimental data from the ALICE collaboration~\cite{ALICE:2023yuk} for $R_T$ integrated events are shown as the open markers for comparison. The AMPT model reasonably reproduces the $p_T$ spectra of associate particles across all particle species up to 3 GeV$/c$, although a  slight underestimation of $K$ production at high $p_T$ is observed. The characteristic mass dependent shift of the peaks in $p_T$ spectra from pion to proton has been captured by the AMPT calculation. It is also observed in this comparison that the transverse region is more sensitive to the variations of event activity changing from $0<R_T<0.5$ to $2.5<R_T<5$. The $p_T$ distribution in transverse region with $2.5<R_T<5$ is harder than the low $R_T$ case and becomes similar to the shape of $p_T$ spectra in the toward region. 
   
   To further explore the difference between the $p_T$ spectra in toward and transverse region, we present the ratio of the spectra between these two topological regions within different $R_T$ event classes in Fig.~\ref{fig:tow_to_trans_ratio}.  The ratio evolves from an increasing trend at low $R_T$ to an almost flat behavior at high $R_T$. It is shown in this comparison that the spectra in the toward region are noticeably harder than those in the transverse region especially in the low activity events, reflecting the strong influence of jet fragmentation in the toward region.  The transverse region spectra are predominantly shaped by MPI process at low to intermediate $R_T$, leading to a rapid increase of the average $p_T$. At high $R_T$, the hardening of $p_T$ spectra might arise from the increasing gluon emissions within a jet, indicating that jet effects become increasingly important in the transverse region when the overall event activity is sufficiently high.

    \begin{figure*}[htbp!]
		\begin{center}
			\includegraphics[width=0.95\textwidth]{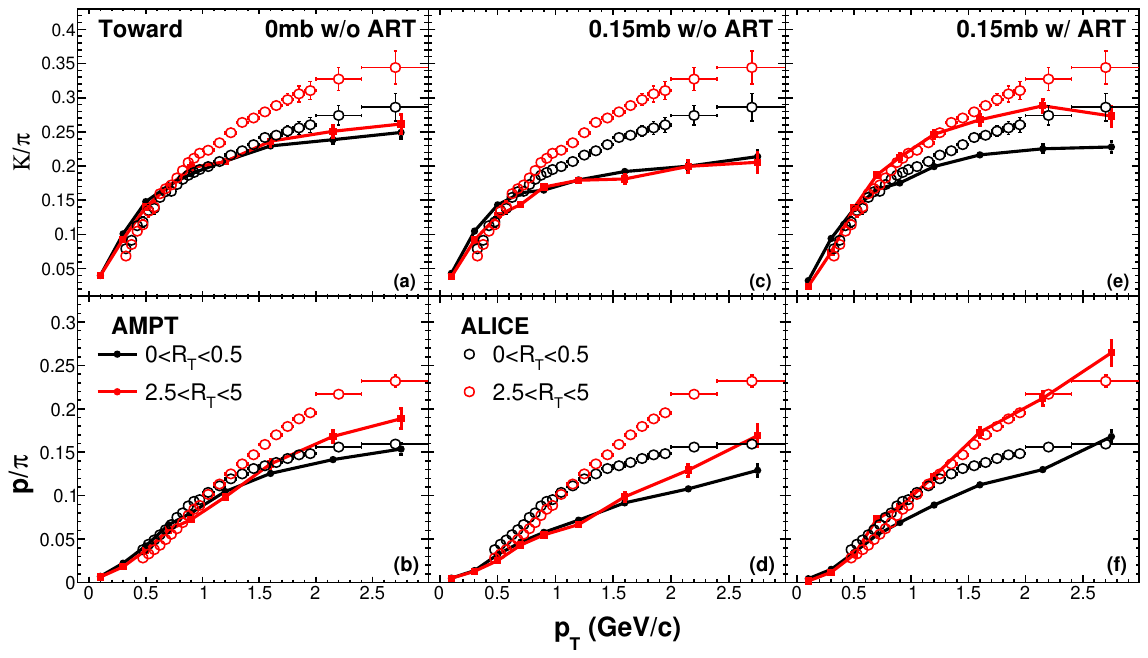}
			\includegraphics[width=0.95\textwidth]{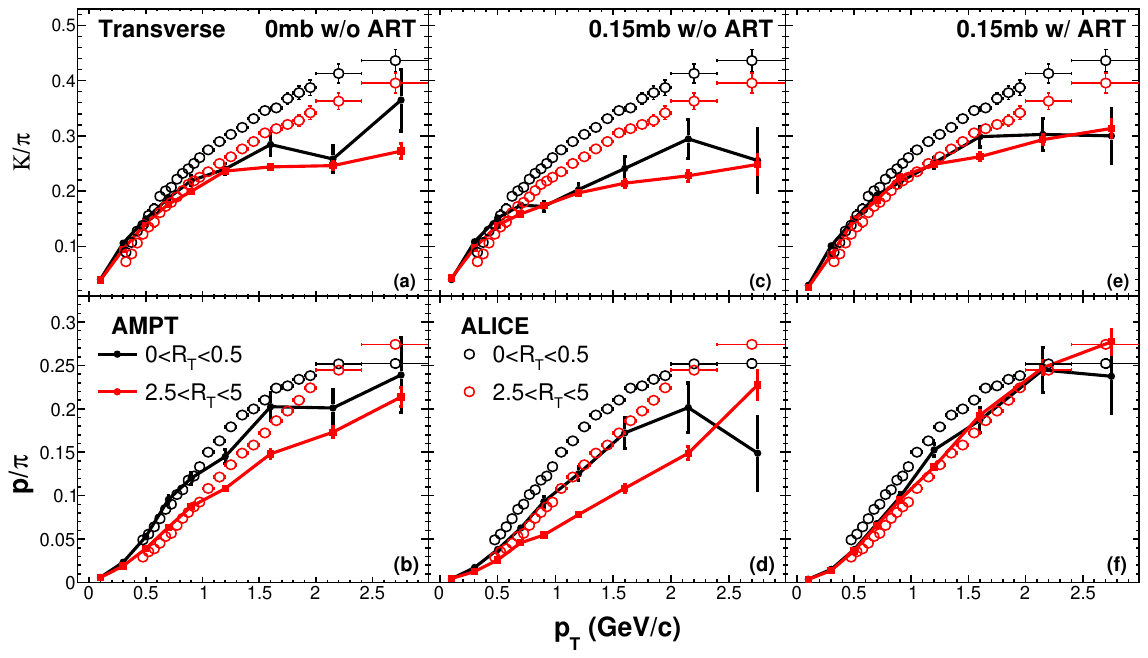}
			\caption{K/$\pi$ and p/$\pi$ ratios as a function of $p_T$ for two $R_T$ intervals in toward (upper panel) and transverse (lower panel) region. 	The AMPT results without final state rescatterings (0 mb w/o ART), with only final state parton rescatterings (0.15 mb w/o ART), with all final state interactions (0.15 mb w/ ART) are shown with black and red lines in left, middle and right column, respectively. ALICE data within corresponding $R_T$ ranges taken from Ref.~\cite{ALICE:2023yuk} are indicated by the open markers for comparison.} 
			\label{fig:ratio_pt_tow}
		\end{center}
	\end{figure*}

    \begin{figure*}[htbp!]
	\begin{center}
		\includegraphics[width=1\textwidth]{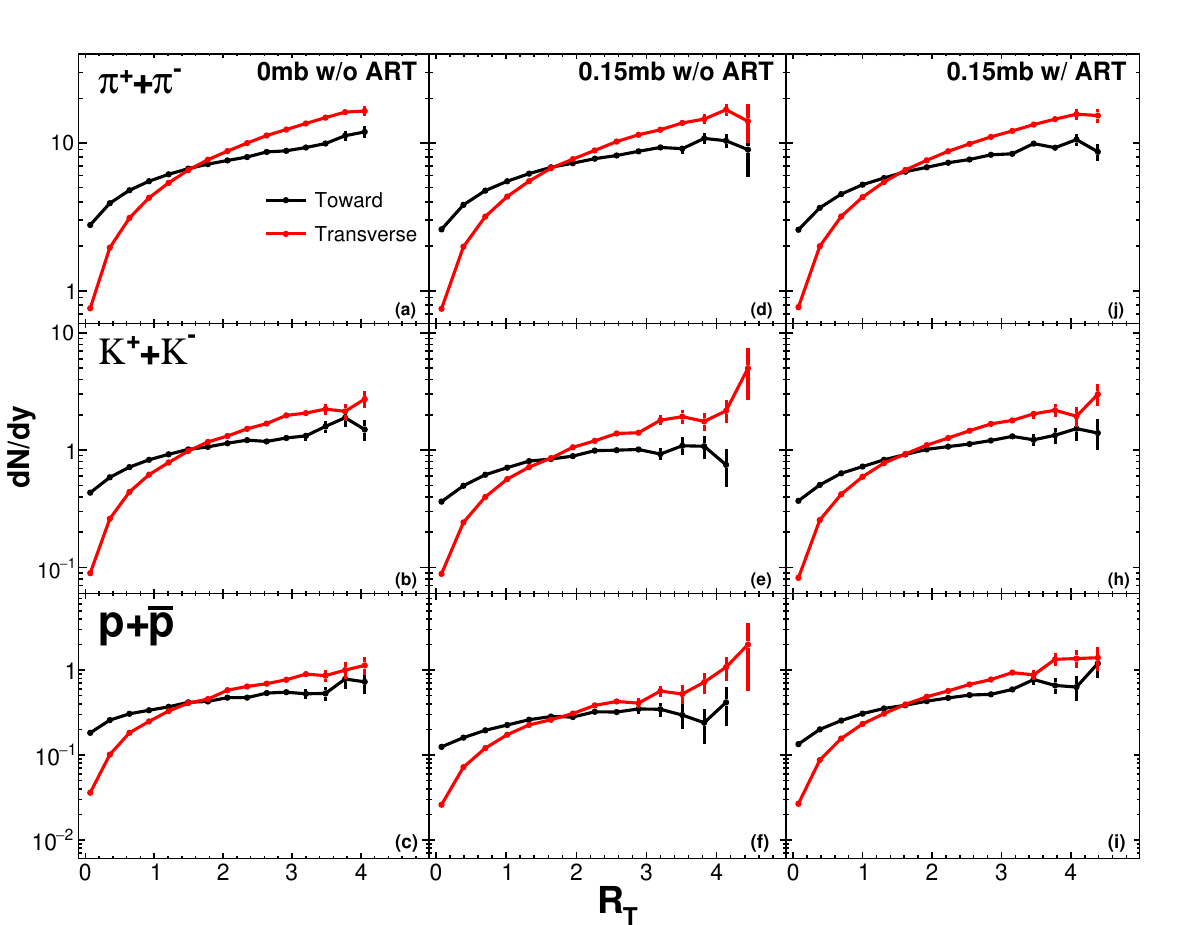}
		\caption{Mid-rapidity particle yields as a function of $R_T$ for $\pi$ (top row), $K$ (middle row) and proton (bottom row). The results in the toward and transverse regions are shown black and red lines. The AMPT calculations without final state rescatterings (0 mb w/o ART), with only final state parton rescatterings (0.15 mb w/o ART), with all final state interactions (0.15 mb w/ ART) are shown in left, middle and right column, respectively. 
		} \label{fig:yield_Rt}
	\end{center}
\end{figure*}

    In Fig.~\ref{fig:ratio_pt_tow}, we further examine the $p_T$ differential K/$\pi$ and p/$\pi$ ratios from AMPT model in two $R_T$ intervals: $0\leq R_T<0.5$ (black line) and $2.5<R_T<5$ (red line). The toward and transverse region results are presented in the upper and lower panels, respectively. With the event activity leverage, the $p_T$ differential particle ratios in toward and transverse region can be used to uncover the sensitivity of different event topological regions to the medium effects more systematically. To quantify the impact of the final state evolution, the partonic and hadronic cascade effects within the AMPT model are enabled sequentially. The left column displays results with all final state interactions turned off, the middle column includes only partonic cascade effects, and the right column incorporates both partonic and hadronic final state interactions. This approach allows us to disentangle the contributions of different stages of the medium evolution to the observed particle ratios. Experimental data from the ALICE collaboration~\cite{ALICE:2023yuk} in two $R_T$ event classes are shown as the open markers with corresponding color coding for comparison.
    
    As indicated in by Fig.~\ref{fig:ratio_pt_tow}, particle ratios in the toward region from AMPT show negligible dependence on $R_T$ when final state interactions are turned off. It is observed that the inclusion of partonic interactions and the corresponding coalescence process decreases the particle ratios. This modification stems from our choice of coalescence parameters, optimized to describe the final inclusive baryon production, as detailed in our previous work~\cite{Zhang:2025pqu}.  In particular, the production of light flavor baryons is suppressed relative to the predictions from string fragmentation, while the enhanced total strange baryon yield influences the production of strange mesons. It is also noticed that the parton stage evolution effects are already significant in lower event activity events and appear to show weaker $R_T$ dependence in the toward region for higher event activity. This behavior can be understood in the context that partonic rescatterings are sensitive to the high local parton density near the leading jet in the toward region. Even in the events with low overall UE activity, the local density in the toward region is high enough to induce significant partonic rescatterings. With increasing $R_T$, the overall UE activity and global density rises, but the partonic rescattering effects in the toward region shows a weaker dependence because the local density near the leading jet is already high and reaches a plateau.
    In contrast, the hadronic cascade effects usually become more important when the event activity is higher, contributing significantly to the observed $R_T$ related enhancement for both K/$\pi$ and p/$\pi$ at higher $p_T$. The hadronic rescattering stage begins after the partonic interactions have ceased and therefore occurs in a more dilute system. Its impact becomes substantial only when the global density is sufficiently high, as in events with large $R_T$. It is noted that such substantial hadronic effects on hard parton fragmentation have also been reported in hybrid hadronization frameworks, extending even to dilute collision systems~\cite{Roch:2025tbx,Dorau:2019ozd}. 
    
    As is shown in the lower panel of Fig.~\ref{fig:ratio_pt_tow}, the K/$\pi$ ratio in the transverse region does not show strong $R_T$ dependence across the most $p_T$ region, irrespective of whether final state interactions are included. However, the transverse region p/$\pi$ ratio in high $R_T$ event class is found to be lower than that in the low $R_T$ events when all the final state interactions are turned off as shown by the left column of Fig.~\ref{fig:ratio_pt_tow}. This behavior is related to the MPI initiated inter-string reconnection effects in the UE process. It is also interesting to see this string fragmentation driven $R_T$ dependence effects are quite weak in the toward region, implying the sizable $R_T$ dependence in toward region a unique feature of final state medium evolution effects. The partonic stage rescattering process further enlarges the difference between two $R_T$ event classes, as an outcome of the growing global density in the transverse region associated with the event activity variations. The p/$\pi$ ratio in high $R_T$ events is significantly suppressed in the partonic evolution stage due to the kinematic shifts of baryon objects in the coalescence process. The follow up hadronic rescattering included in the right column of this figure enhances the p/$\pi$ ratio and narrow down the difference between the two $R_T$ event classes as opposed to the parton stage evolution effects.     
    It is shown in this comparison that including both final state parton and hadron cascade effects generally improve the description to the K/$\pi$ and p/$\pi$ experimental data both in toward and transverse region compared to the case without final state interactions in the AMPT model, suggesting the final state medium evolution effects can be important ingredients to understand the event activity dependent particle ratios measured in high energy pp collisions at the LHC energy.

	Our model calculation, shown by Fig.~\ref{fig:yield_Rt}, implies that the $p_T$ integrated yields are increasing with $R_T$ in both toward and transverse regions for all particle species in a similar way. However, the increase is much more pronounced in the transverse region, while the toward region yield exhibits only a mild rise. This difference largely originates from the intrinsic autocorrelation effect, since $R_T$ is defined using the multiplicity of particles in the transverse region. At low $R_T$, the yields in both regions grow rapidly, dominated by MPI that enhance the soft underlying event. In contrast, the increase at high $R_T$ becomes milder and is mainly governed by jet fragmentation, which contribute additional particles without substantially altering the global event multiplicity. The toward region hadron yields are higher than the transverse region at low $R_T$ but become lower at $R_T$ close to 1.5. The crossing point between toward and transverse region is almost the same for all particle species and universal for different final state evolution configurations, suggesting a transition from MPI driven to jet dominated physics.
	As the toward region is convoluted by the jet and UE contribution effects, we disentangle the jet behavior by removing the UE components using the toward subtracting transverse method detailed in Sect.~\ref{sec:rt_def}. This approach allows us to study in-jet production of identified hadrons. Moreover, it is noted that both the AMPT model calculations and the experimental data indicate that the in-jet hadron yield defined by the difference between toward with respect to the transverse yield becomes negative especially when $R_T$ is high. The effective $R_T$ range used for this in-jet subtraction analysis is determined from the behavior of the yields shown in Fig.~\ref{fig:yield_Rt}.

    \begin{figure*}[htbp!]
	\begin{center}
		\includegraphics[width=\textwidth]{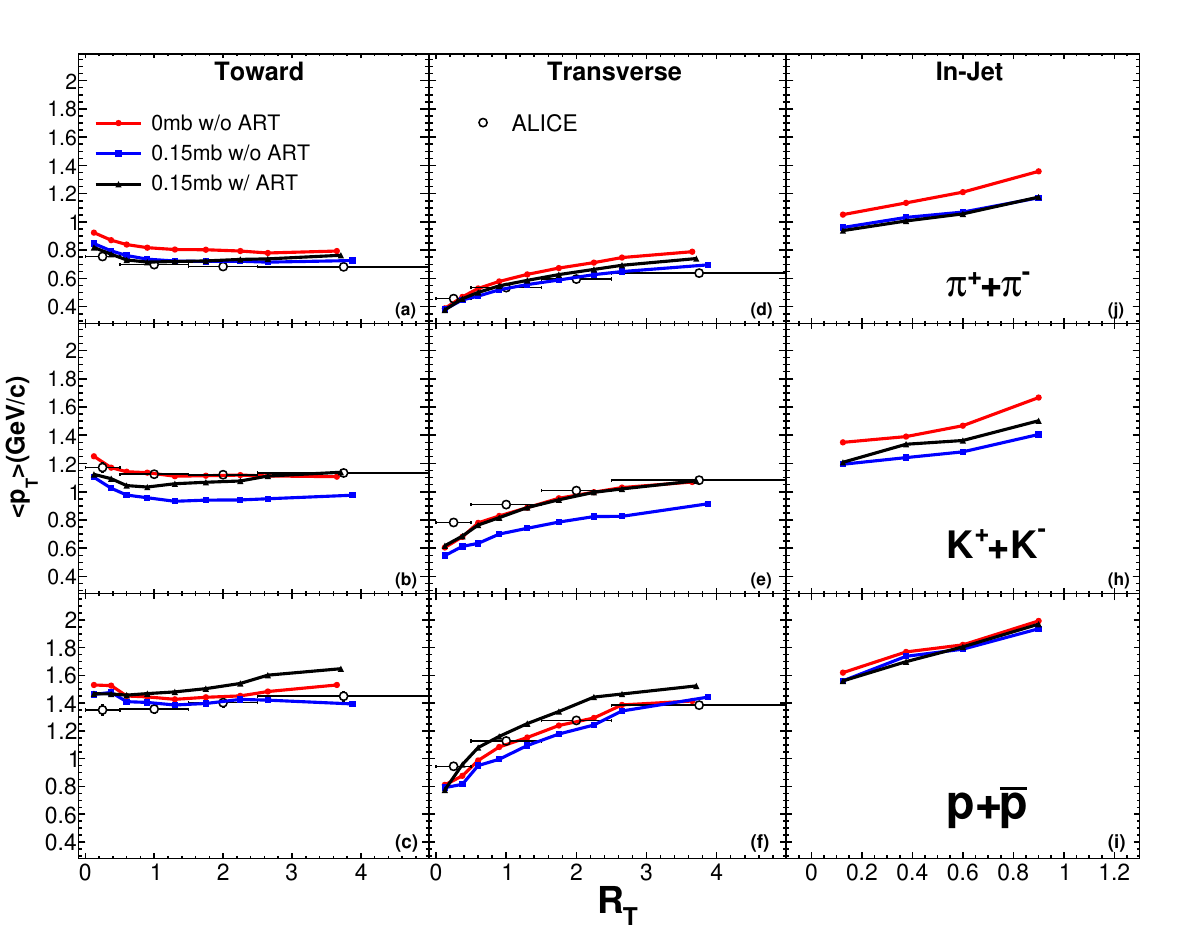}	
		\caption{Average $p_T$ for $\pi$, K and p from toward (left column), transverse (middle column) and in-jet (right column) hadron productions as a function of $R_T$. The AMPT results without final state rescatterings (0 mb w/o ART), with only final state parton rescatterings (0.15 mb w/o ART), with all final state interactions (0.15 mb w/ ART) are shown with red, blue and black lines, respectively.ALICE data for toward and transverse region results taken from Ref.~\cite{ALICE:2023yuk} are indicated by the open markers for comparison.} 
		\label{fig:AvgPt}
	\end{center}
	\end{figure*}    
	In Fig.~\ref{fig:AvgPt}, we explore the $R_T$ dependence of average $p_T$ for pion, kaon and proton within toward (left column), transverse (middle column) and in-jet (right column) topological regions with and without final state interactions in AMPT.  It is observed in this comparison that both toward and transverse region average $p_T$ are subject to the final state interaction effects in AMPT. When final state partonic interactions are included, the average $p_T$ becomes smaller for all particle species, indicating a sizable parton level energy loss effects. The final state hadronic interactions will change the magnitude of $p_T$ for kaon and proton significantly while pion is less affected. The enhancement of proton and kaon $\langle p_T \rangle$ in the hadronic stage can be understood microscopically through the pion wind effect~\cite{De:2022yxq,Elfner:2022iae}. During the hadronic cascade stage, the system is dominated by a dense pion gas that expands outward. Through frequent hadronic scatterings, heavier particles such as kaons and protons are dragged along by this co-moving pion gas. The collective outward motion of the pion medium transfers transverse momentum to heavier hadrons, effectively boosting their average $p_T$.
	The impact on proton is quite strong and has a non-trivial $R_T$ dependence which eventually leads to an increasing behavior of average $p_T$ for proton in high activity events in toward region, which can be connected to the previous particle ratio features in Fig.~\ref{fig:ratio_pt_tow}. It is also found that the AMPT model including final state partonic and hadronic evolution effects generally agree with data reasonably, although it slightly overestimates proton $\langle p_T \rangle$ at high $R_T$. The in-jet average $p_T$ is usually significantly larger than that observed in toward and transverse region, indicating strong jet fragmentation effects. It is also shown that the in-jet pion and kaon average $p_T$ shows similar final state energy loss behavior in AMPT found in toward and transverse region while the proton in-jet average $p_T$ is less sensitive to the final state effects. This difference might come from the fact that the energy loss modification is mainly restricted to the low $p_T$ region and the proton average $p_T$ is sufficiently large to be less impacted by these variations.

    \begin{figure*}[htbp!]
		\begin{center}
			\includegraphics[width=\textwidth]{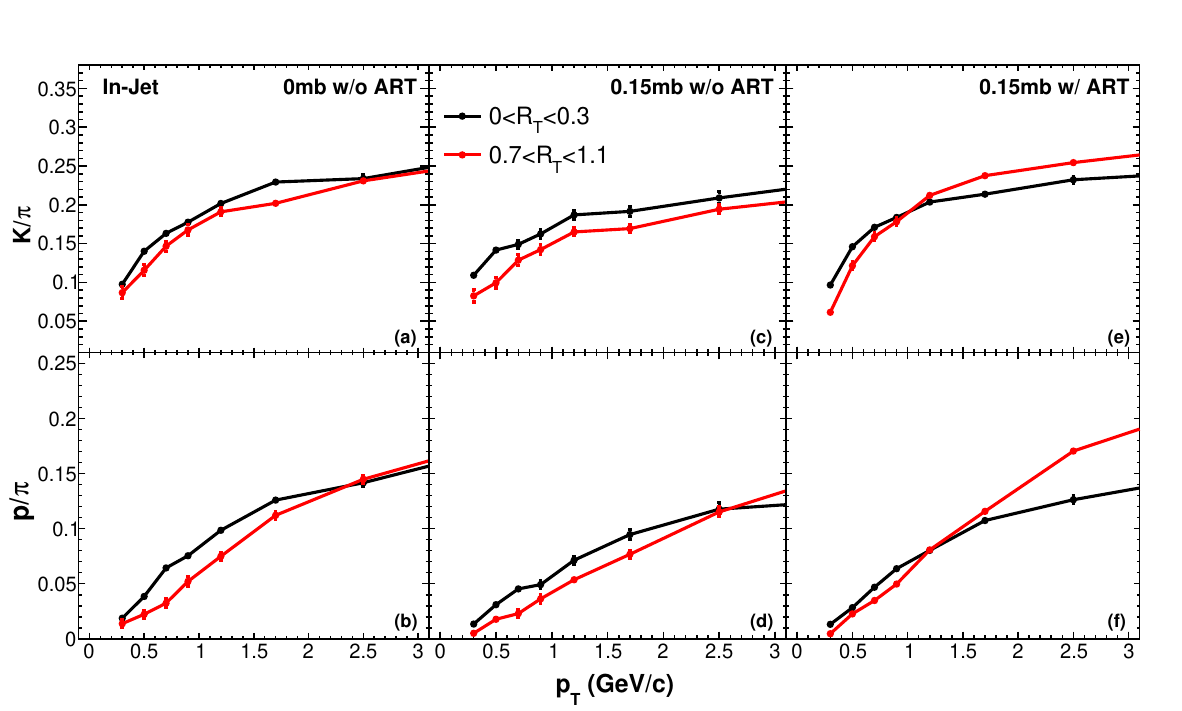}
			\caption{K/$\pi$ and p/$\pi$ ratios of in-jet hadron productions as a function of $p_T$ for two $R_T$ intervals. The AMPT results without final state rescatterings (0 mb w/o ART), with only final state parton rescatterings (0.15 mb w/o ART), with all final state interactions (0.15 mb w/ ART) are shown with black and red lines in left, middle and right column, respectively.} 
			\label{fig:injet_ratio_Pt}
		\end{center}
	\end{figure*}

    Figure.~\ref{fig:injet_ratio_Pt} shows the $p_T$ dependence of in-jet K/$\pi$ ratio and p/$\pi$ ratio in two $R_T$ event classes with different final state interaction configurations in the AMPT model. The in-jet observables are derived from a subtraction of two large yields, which particularly in high $R_T$ events can result in negative yields at low $p_T$. The $R_T$ bins chosen for this figure, $0<R_T<0.3$ and $0.7<R_T<1.1$, were selected to ensure that the subtracted observable remains positive over the entire $p_T$ range, while still representing distinct low- and high-activity event classes. The in-jet particle ratios are generally smaller than those in the toward region, consistent with the expectation that particle production dominated by genuine jet fragmentation yields smaller baryon-to-meson and strangeness-to-pion ratios compared to UE contributions. This observation supports the effectiveness of the subtraction method in isolating purified jet components from UE backgrounds. It is shown in the left column that the both K/$\pi$ and p/$\pi$ ratio become smaller as $R_T$ increases across almost the entire $p_T$ range when no final state interactions are incorporated. This behavior can be interpreted as a result of depletion of gluon radiations with the change of event activity in the leading jet. Requiring larger event activity makes the leading jet energy redistribute to the UE region and limits the phase space for gluon radiation of the leading jet, thus reduces the string tension associated with the leading jet during string fragmentation process and delivers fewer strange quarks and diquarks as the string breaks up~\cite{Zheng:2018yxq}. Moreover, at higher $p_T$, the in-jet particle ratios with different event activity tend to converge, reflecting the universality of jet fragmentation at high momenta. Similar behavior has also been reported in previous jet reconstruction studies~\cite{ALICE:2022ecr}. 
    When partonic final-state interactions are included, as shown in the middle column, the overall particle ratios are further suppressed, suggesting that parton-level rescatterings can modify the hadron composition even in jet-associated regions, consistent with the findings in Fig.~\ref{fig:ratio_pt_tow}. A crossing point between the different $R_T$ event classes begins to emerge in the p/$\pi$ ratio due to the coalescence hadronization process. With the inclusion of hadronic rescattering, as shown in the right column, the enhancement of K/$\pi$ and p/$\pi$ ratios at intermediate and high $p_T$ becomes more pronounced for high $R_T$ events. This leads to a distinct crossing behavior in the particle ratios between high and low $R_T$ event classes at intermediate $p_T$ driven by late stage hadronic interactions such as the pion wind effect. This crossing behavior, progressively amplified through partonic and hadronic final state interactions, serves as a sensitive probe of jet medium interactions in high multiplicity small systems. It highlights the combined influence of parton energy loss, coalescence, and hadronic rescattering in shaping the particle composition within jets, providing possible evidence for medium induced modifications and partial energy loss effects in small collision systems. We also notice that varying the partonic cross section within a reasonable range does not change the qualitative trends, indicating that our main conclusions are robust against uncertainties in the partonic rescattering cross section parameter.

\section{Summary}
\label{sec:summary}
	
In this study, we employed the AMPT model with PYTHIA8 initial conditions to investigate the production of pion, kaon and proton in pp collisions at $\sqrt{s} = 13$ TeV with a focus on disentangling jet fragmentation effects from the underlying event process contributions. By classifying events according to the relative transverse activity, we explore the $R_T$ dependence of the $p_T$ distribution for identified hadron species and particle ratios in the toward and transverse region. Our results demonstrate that the AMPT model with both partonic and hadronic final state rescattering effects can satisfactorily describe the experimental data in the UE dominated transverse region as well as the jet aligned toward region. We find that the hadronic final state interactions are important to understand the $R_T$ dependent splitting of particle to pion ratios at $p_T$ around 2 GeV$/c$ in the toward region. To isolate the jet related modifications more exclusively, we construct the in-jet hadron yield by taking the difference between the toward and transverse region hadron productions. It is shown that the final state interactions in AMPT will lead to a non-trivial crossing of in-jet p/$\pi$ ratios at intermediate $p_T$ between low $R_T$ and high $R_T$ event classes. This behavior can be identified as a sensitive probe of jet medium interactions in high multiplicity pp events to explore the modification of the jet's chemistry and fragmentation pattern, providing a novel observable to examine the effects of jet energy loss in high activity small systems. This technique provides a complementary tool to full jet reconstruction, offering enhanced sensitivity to the modification of low $p_T$ jet fragments where traditional reconstruction is challenging. Future extension of this  methodology to larger collision systems and systematic comparison with the full jet reconstruction analysis can be of great interest to map the evolution of jet quenching phenomena from small to large systems.

	\begin{acknowledgments}
This work was supported by the National Key Research and Development Program of China (Grant No. 2024YFA1610800) and the National Natural Science Foundation of China (Nos. 12205259, 12147101, 12275103, 12061141008) and the Fundamental Research Funds for the Central Universities, China University of Geosciences(Wuhan) with No. G1323523064 and the Innovation Fund of Key Laboratory of Quark and Lepton Physics QLPL2025P01.
		
	\end{acknowledgments}


	\bibliography{ref_injet}

\end{document}